\documentstyle[proceedings,numreferences]{crckapb}
\begin{opening}

\title{Cauchy-characteristic matching  }
\author{Nigel T. Bishop}
\institute{Department of Mathematics, Applied Mathematics and Astronomy\\
      University of South Africa, P.O. Box 392, Pretoria 0003, South Africa}
\author{Roberto Gomez}
\author{Luis Lehner}
\author{Bela Szilagyi}
\author{Jeffrey Winicour}
\institute{Department of Physics and Astronomy\\
      University of Pittsburgh, Pittsburgh, PA 15260}
\author{Richard A. Isaacson}
\institute{Department of Physics and Astronomy,\\
	University of Pittsburgh, Pittsburgh, PA 15260\\
	\hspace*{2.5cm} and\\
	Physics Division, National Science Foundation\\
	4201 Wilson Blvd., Arlington VA 22230.}

\end{opening}
\begin{document}

\begin{abstract}

This paper gives a detailed pedagogic presentation of the central
concepts underlying a new algorithm for the numerical solution of
Einstein's equations for gravitation. This approach incorporates the
best features of the two leading approaches to computational
gravitation, carving up spacetime via Cauchy hypersurfaces within a
central worldtube, and using characteristic hypersurfaces  in its
exterior to connect this region with null infinity and study
gravitational radiation. It has worked well in simplified test
problems, and is currently being used to build computer codes to
simulate black hole collisions in 3-D.

\end{abstract}

\section{Preamble}

Throughout his career in theoretical gravitational physics, Vishu has
been interested in questions of black holes and gravitational radiation.
The subject of his useful early study of radiation from binary
systems~\cite{vishubinary} is now at the forefront of research as the
principal target for the first exciting experimental measurements of the
LIGO project. His fundamental studies of the properties of black hole
excitations and radiation~\cite{vishuRW} used the formidable technology
of the mid 1960's, i.e. Regge-Wheeler perturbation theory, to
analytically extract the crucial physical result that black holes were
stable. In his continuing studies of the interaction of black holes with
gravitational radiation~\cite{vishuscattering}, he first demonstrated
the phenomenon that was later to be called normal mode excitations of
black holes, and noted that the frequency of the emitted radiation
carried with it key information which could be used to determine the
mass of the invisible black hole. These issues are still at the heart of
current research three decades later. 

Today, the technology to study these questions has become even more
formidable, requiring large groups of researchers to work hard at
developing computer codes to run on the massively parallel
supercomputers of today and tomorrow. This has caused gravitational
theory to enter into the realm of ``big science" already familiar to
experimental physics, with large, geographically distributed
collaborations of scientists engaged on work on expensive, remote,
central facilities. The goal of this modern work is to understand
the full details of black hole collisions, the fundamental two-body
problem for this field. Recent developments in this area are very
encouraging, but it may well take another three decades until all the
riches of this subject are mined. 

In this paper, we will present all the gory details of how the best
current methods in computational gravitation can be forged into a single
tool to attack this crucial problem, one which is currently beyond our
grasp, but perhaps not out of our reach.

\section{Introduction}

Although Einstein's field equations for gravitation have been known for
the past 80 years, their complexity has frustrated attempts to extract
the deep intellectual content hidden beneath intractable mathematics.
The only tool with potential for the study of the general
dynamics of time-dependent, strongly nonlinear gravitational fields
appears to be computer simulation. Over the past two decades, two
alternate approaches to formulating the specification and evolution of
initial data for complex physical problems have emerged. The {\it
Cauchy} (also known as the ADM or $``3+1"$) approach foliates spacetime
with spacelike hypersurfaces. Alternatively, the {\it characteristic}
approach uses a foliation of null hypersurfaces. 

Each scheme has its own different and complementary strengths and
weaknesses. Cauchy evolution is more highly developed and has
demonstrated good ability to handle relativistic matter and strong
fields. However, it is limited to use in a finite region of spacetime,
and so it introduces an outer boundary where an artificial boundary
condition must be specified. Characteristic evolution allows the
compactification of the entire spacetime, and the incorporation of
future null infinity within a finite computational grid. However, in
turn, it suffers from complications due to gravitational fields causing
focusing of the light rays. The resultant caustics of the null cones
lead to coordinate singularities. At present, the unification of both
of these methods~\cite{nb93} appears to offer the best chance for
attacking the fundamental two-body problem of modern theoretical
gravitation: the collision of two black holes.

The basic methodology of the new computational approach called {\it
Cauchy-characteristic matching (CCM)}, utilizes Cauchy evolution within
a prescribed world-tube, but replaces the need for an outer boundary
condition by matching onto a characteristic evolution in the exterior to
this world-tube, reaching all the way out to future null infinity. The
advantages of this approach are: (1) Accurate waveform and polarization
properties can be computed at null infinity; (2) Elimination of the
unphysical outgoing radiation condition as an outer boundary condition
on the Cauchy problem, and with it all accompanying contamination from
spurious back-reflections, consequently helping to clarify the Cauchy
initial value problem. Instead, the matching approach incorporates
exactly all physical backscattering from true nonlinearities; (3)
Production of a global solution for the spacetime; (4) Computational
efficiency in terms of both the grid domain and algorithm. A detailed
assessment of these advantages is given in Sec.~\ref{sec:advan}.

The main modules of the matching algorithm are:

\begin{itemize}

\item The outer boundary module which sets the grid structures.

\item The extraction module whose input is Cauchy grid data in the
neighborhood of the world-tube and whose output is the inner boundary
data for the exterior characteristic evolution.

\item The injection module which completes the interface by using the
exterior characteristic evolution to supply the outer Cauchy boundary
condition, so that no artificial boundary condition is necessary.

\end{itemize}

Details of the Cauchy and characteristic codes have been presented
elsewhere. In this paper, we present only those features
necessary to discuss the matching problem.

\section{Advantages of Cauchy-characteristic matching (CCM)}
\label{sec:advan}

There are a number of places where errors can arise in a pure Cauchy
computation. The key advantage of CCM is that there is tight control
over the errors, which leads to computational efficiency in the
following sense. For a given target error $\varepsilon$, what is the
amount of computation required for CCM (denoted by $A_{CCM}$) compared
to that required for a pure Cauchy calculation (denoted by $A_{WE}$)? It
will be shown that $A_{CCM}/A_{WE} \rightarrow O$ as $\varepsilon
\rightarrow O$, so that in the limit of high accuracy CCM is by far the
most efficient method.

In CCM a ``3 + 1'' interior Cauchy evolution is matched to an exterior
characteristic evolution at a world-tube of constant radius $R$. The important
point is that the characteristic evolution can be rigorously compactified, so
that the whole spacetime to future null infinity may be represented on a
finite grid.  From a numerical point of view this means that the only error
made in a calculation of the gravitational radiation at infinity is that due
to the finite discretization $h$;  for second-order algorithms this error is
$O(h^2)$.
The value of the matching radius $R$ is important, and it will turn out
that for efficiency it
should be as small as possible.  The difficulty is that if $R$ is too small
then caustics may form.  Note however that the
smallest value of $R$ that avoids caustics is determined by the physics of
the problem, and is {\em not} affected by either the discretization
$h$ or the numerical method.

On the other hand, the standard approach is to make an estimate of the
gravitational radiation solely from the data calculated in a pure Cauchy
evolution. The simplest method would be to use the raw data, but that
approach is too crude because it mixes gauge effects with the physics. Thus
a substantial amount of work has gone into methods to factor out the gauge
effects and to produce an estimate of the gravitational field at null
infinity from its behavior within the domain of the Cauchy
computation~\cite{ab1,ab2,ab3}. We will call this method {\it waveform
extraction, or WE}. The computation is performed in a domain $D$, whose
spatial cross-section is finite and is normally spherical or cubic. Waveform
extraction is computed on a world-tube $\Gamma$, which is strictly in the
interior of $D$, and which has a spatial cross-section that is spherical and
of radius $r_E$. While WE is a substantial improvement on the crude approach,
it has limitations. Firstly, it disregards the effect, between $\Gamma$ and
null infinity, of the nonlinear terms in the Einstein equations; the
resulting error will be estimated below. Secondly, there is an error, that
appears as spurious wave reflections, due to the inexact boundary condition
that has to be imposed at $\partial D$. However, we do not estimate this
error because it is difficult to do so for the general case; and also
because it is in principle possible to avoid it by using an exact
artificial boundary condition (at a significant computational cost).

The key difference between CCM and WE is in the treatment of the
nonlinear terms between $\Gamma$ and future null infinity. WE ignores these
terms, and this is an inherent limitation of a perturbative method (even if
it is possible to extend WE beyond linear order, there would necessarily
be a cut-off at some finite order).
Thus our strategy for comparing the computational efficiency of CCM and WE
will be to find the error introduced into WE from ignoring the
nonlinear terms; and then to find the amount of computation needed to control
this error.

\subsection{Error estimate in WE}

As discussed earlier, ignoring the nonlinear terms between $\Gamma$ (at
$r=r_E$) and null infinity introduces an error, which we estimate using
characteristic methods. The Bondi-Sachs metric is
\begin{eqnarray}
   ds^2 &=&-\left(e^{2\beta}{V \over r}
	       -r^2h_{AB}U^AU^B\right)du^2 \nonumber \\
        & & \mbox{} -2e^{2\beta}dudr
	-2r^2 h_{AB}U^Bdudx^A
	+r^2h_{AB}dx^Adx^B,    \label{eq:bmet}
\end{eqnarray}
where $A, B = 2,3$ and $h_{AB}$ is a spherical metric that is completely
described by one complex function $J$. The initial data required on a null
cone $u~=constant$ is $J$, and the hypersurface equations $R_{1\alpha}=0$
then form a hierarchy from which $\beta$, $U^A$ and $W\equiv(V-r)/r^2$ are found~\cite{hpn}.
The evolution equation $R_{AB}h^{AB}=0$~\cite{hpn} is
\begin{equation}
2(rJ)_{,ur} = L_J + N_J
\label{eq:WE1}
\end{equation}
where $L_J$ represents the linear part and $N_J$ the nonlinear part.

The order of magnitude of various terms can be expressed in terms of a
function $c(u,x^A)$ (whose time derivative $c_{,u}$ is the news function);
note that $c$ is not a small quantity. The expressions are
\begin{equation}
J=O(\frac{c}{r}),
\beta=O(\frac{c^2}{r^2}),
U^A=O(\frac{c}{r^2}),
W=O(\frac{c^2}{r^2}).
\label{eq:WE2}
\end{equation}
These estimates are obtained from the hypersurface equations,
and assume that
the background geometry is Minkowskian.  Should this not be the case then
constants of order unity would be added, and the effect of this
would be to amend (\ref{eq:WE1}) by adding terms to $L_J$ so that it
represents wave propagation on a non-Minkowskian background. However, the
order of magnitude of terms in $N_J$ would not be affected.
It is straight forward to confirm that $N_J$ involves terms of order
\begin{equation}
O(\frac{c^2}{r^3}).
\label{eq:WE3}
\end{equation}
WE estimates the news at future null infinity from data at $r=r_E$, and
could be made exact if $N_J$ were zero. Thus the error introduced by
ignoring $N_J$ is
\begin{equation}
\varepsilon(c_{,u}) \equiv (c_{,u})_{exact}-(c_{,u})_{WE} =
\int_{r_E}^{\infty}O(\frac{c^2}{r^3}) dr = O(\frac{c^2}{r_E^2}).
\label{eq:WE4}
\end{equation}

This would be the unavoidable linearization error in WE were it
implemented as an ``exact'' artificial boundary condition
by using global techniques, such as the difference potential method, to
eliminate back reflection at the boundary~\cite{exabc}. However, this is
computationally expensive~\cite{eng} and has not even been attempted in
general relativity. The performance of WE continues to improve but
the additional error due to back reflection remains~\cite{alli}.

\subsection{Computational efficiency}

A numerical calculation of the emission of gravitational radiation
using a CCM algorithm is expected to be second-order convergent, so
that after a fixed time interval the error is:
\begin{equation}
\varepsilon = O(h^2) \simeq k_1 h^2,
\label{eq:CE1}
\end{equation}
where $h$ is the discretization length.  On the other hand, the same
calculation using WE must allow for the error found in (\ref{eq:WE4}), and
therefore after the same fixed time interval there will be an error of:
\begin{equation}
\varepsilon=O(h^2,r_E^{-2}) \simeq max(k_2h^2,\frac{k_3}{r_E^2}).
\label{eq:CE2}
\end{equation}

We now estimate the amount of computation required for a given desired
accuracy.  We make one important assumption:
\begin{itemize}
\item The computation involved in matching, and in waveform extraction,
is an order of magnitude smaller than the computation involved in
evolution, and is ignored.
\end{itemize}
For the sake of transparency we also make some simplifying assumptions;
if not true there would be some extra constants of order unity in the formulas
below, but the qualitative conclusions would not be affected.
\begin{enumerate}
\item The amount of computation per grid point per time-step, $a$, is the same
for the Cauchy and characteristic algorithms.
\item  The constants $k_1, k_2$ in the equations above are approximately
equal and will be written as $k$.
\item In CCM, the numbers of Cauchy and characteristic grid-points
are the same; thus the total number of grid points per time-step is:
\begin{equation}
     \frac{8 \pi R^3}{3 h^3}.
\label{eq:CE3}
\end{equation}
\item In WE, the number of grid points in $D$ is twice the number contained
in $\Gamma$; thus the total number of grid points per time-step is
\begin{equation}
   \frac{8 \pi r_E^3}{3 h^3}.
\label{eq:CE4}
\end{equation}
\end{enumerate}
It follows that the total amount of computation $A$ (i.e., number of
floating-point operations) required for the two methods is:
\begin{equation}
A_{CCM}=\frac{8 \pi R^3 a}{3 h^4},\ \ A_{WE}=
\frac{8 \pi r_E^3 a}{3 h^4}.
\label{eq:CE5}
\end{equation}
Thus $R > r_E$ or $R < r_E$ determines which method requires the least
amount of computation. Because of the assumptions (1) to (4) this criterion
is not exact but only approximate.

As stated earlier, in a given physical situation the minimum allowed
value of $R$ is determined by the physics. However, $r_E$ is
determined by the target error (equation (\ref{eq:CE2})); and there is
also a minimum value determined by the
condition that the nonlinearities must be sufficiently weak for a
perturbative expansion to be possible.  Thus, in a loose
sense, the minimum value of $r_E$ is expected to be related to the minimum value
of $R$. It follows that the computational efficiency of a CCM algorithm
is never  expected to be significantly worse than that of a WE algorithm.

If high accuracy is required, the need for computational efficiency always
favors CCM.  More precisely, for a given desired error
$\varepsilon$, equations (\ref{eq:CE1}) and (\ref{eq:CE2}), and
assumption (2), imply:
\begin{equation}
h=\sqrt{\varepsilon/k},\ r_E=\sqrt{k_3/\varepsilon}.
\label{eq:CE6}
\end{equation}
Thus, substituting equation (\ref{eq:CE6}) into equation (\ref{eq:CE5}),
\begin{equation}
A_{CCM}=\frac{8 \pi R^3 a k^2}{3 \varepsilon^2},
\ A_{WE}=\frac{8 \pi a k^2 k_3^{3/2}}{3 \varepsilon^{7/2}},
\label{eq:CE7}
\end{equation}
so that
\begin{equation}
\frac{A_{CCM}}{A_{WE}}=\frac{R^3 \varepsilon^{3/2}}{k_3^{3/2}}
\rightarrow 0 \ as \ \varepsilon \rightarrow 0.
\label{eq:CE8}
\end{equation}
This is the crucial result, that the computational intensity of CCM
relative to that of WE goes to zero as the desired error $\varepsilon$
goes to zero.

\section{Extraction}

We describe a procedure by which information about the 4-geometry on
the neighborhood of a world-tube $\Gamma$, obtained during a $3+1$
simulation of Einstein's equations, is used to extract boundary data
appropriate for an exterior characteristic formulation. A numerical
implementation of this characteristic formulation~\cite{weak}, is then
used to propagate the gravitational signal to null infinity, where the
radiation patterns are calculated.
The process we describe here is non-perturbative, and to make it as
portable as possible, it assumes only that the $3+1$ simulation can
provide the $3$-metric, the lapse and the shift by interpolation to a set
of prescribed points.

We start by describing the world-tube $\Gamma$ in
Sec.~\ref{sec:world-tube}, we explain what information needs to be
provided by the $3+1$ simulation around the world-tube in
Sec.~\ref{sec:4dgeometry}, giving details of the transformation to null
coordinates in Sec.~\ref{sec:coordtrans}, and the expression for the
Bondi metric in Sec.~\ref{sec:bondimetric}. This provides the boundary
data needed to start up the characteristic code in
Sec.~\ref{sec:nullbdry}. The characteristic code which takes this
boundary information and calculates the waveforms is described in detail
in~\cite{weak}.

\subsection{Parametrization of the world-tube}\label{sec:world-tube}

The notation and formalism are based on Misner, Thorne and
Wheeler~\cite{mtw}.  Greek indices range from $1$ to $4$, Latin indices
range from $1$ to $3$, and upper case Latin indices refer to
coordinates on the sphere and range from $2$ to $3$.

The intersections $S_{t}$ of the world-tube with the space-like slices
$\Sigma_{t}$ of the Cauchy foliation are topologically spherical, and they can
be parametrized by labels $\tilde y^{A}$, $A=[2,3]$ on the sphere. The
intersections themselves are labeled by the time coordinate of the
Cauchy foliation, $x^{4}=t$. Future oriented null cones emanating from
the world-tube are parametrized by the labels on the sphere $\tilde
y^{A}$ and an affine parameter $\lambda$ along the radial direction,
with $\lambda=0$ on the world-tube.  With the identifications $\tilde
y^{1}=\lambda$ and retarded time $\tilde y^{4} \equiv u=t$, we define a null coordinate system
$\tilde y^{\alpha}=(\tilde{y}^{1}, \tilde{y}^{A}, \tilde{y}^{4})$. We
will later introduce a second null coordinate system
$y^{\alpha}=(y^{1},y^{A},y^{4})$, where $y^{1}=r$, $r$ being a
surface area coordinate, and $y^{A}=\tilde{y}^{A}$, $y^{4}=\tilde{y}^{4}$.

Following~\cite{eth}, we cover the unit sphere with two stereographic
coordinate patches, centered around the North and South poles,
respectively, where the stereographic coordinate is related to the
usual spherical polar coordinates $(\theta,\phi)$ by

\begin{equation}
   \xi_{\rm North} = \sqrt{ {{1 - \cos\theta} 
                            \over {1 + \cos\theta}}} e^{i\phi},
   \quad
   \xi_{\rm South} = \sqrt{ {{1 + \cos\theta} 
                            \over {1 - \cos\theta}}} e^{-i\phi}.
   \label{eq:xi}
\end{equation}
Let $\tilde{y}^{A} \equiv (q,p)$, for $A=[2,3]$, label the points of a
stereographic coordinate patch. We adopt as these labels the real and
imaginary part of the stereographic coordinate, i.e. for
$\xi=\tilde{y}^{2} + i \tilde{y}^{3} = q + i p$.

For the computational implementation of the procedure we
are describing, we introduce a discrete representation of these
coordinate patches

\begin{equation}
   \tilde{y}^{2}_{\tt i} = -1 + ({\tt i}-3) \Delta, 
   \quad
   \tilde{y}^{3}_{{\tt j}} = -1 + ({\tt j}-3) \Delta, 
   \label{eq:starray}
\end{equation}
where the computational spatial grid indices {\tt i,j,k} run from $1$ to
$N$ and $\Delta=2/(N-5)$, $N$ is the stereographic grid size and
$\Delta$ the stereographic grid spacing.

On each patch, we introduce complex null vectors on the sphere
$q^{A}=(P/2)(\delta^{A}_{2}+i\delta^{A}_{3})$, where
$P=1+\xi\bar\xi$. The vectors $q^{A}$, $\bar{q}^{A}$ define a metric on
the unit sphere

\begin{equation}
q_{AB} = \frac{1}{2}(q_{A} \bar q_{B} + \bar q_{A} q_{B}) =
       \frac{4} {P^{2}}
       \left [
              \begin{array}{cc}
              1 & 0
              \\
              0 & 1
              \end{array}
       \right ]
\end{equation}
with determinant $\det(q_{AB})=16/P^4$. The co-vector
$\bar{q}_{A}$ satisfies the orthogonality condition
$\bar{q}_{A}q^{A}=2$, and has components
$q_{A}=(2/P) \left(\delta_{A}^{2} + i \delta_{A}^{3} \right)$.  

Given Cartesian coordinates $x^{i}=(x,y,z)$ on a space-like slice
$\Sigma_{t}$, the intersection $S_{t}$ of the world-tube $\Gamma$ with
$\Sigma_{t}$ is described parametrically by three functions of the
$\tilde{y}^{A}, x^{(0)i} =f^{i}(\tilde{y}^{A})$.
In the following, we will fix the location of the world-tube by
choosing these functions (in stereographic coordinates) as :
\begin{eqnarray}
   f^{x}(\tilde{y}^{A}) & = &
   2 R \left({\displaystyle {{\Re(\xi)} \over {1+\xi\bar\xi}}}\right)
   \nonumber \\
   f^{y}(\tilde{y}^{A}) & = &
   \pm 2 R \left({\displaystyle {{\Im(\xi)} \over {1+\xi\bar\xi}}}\right)
   \nonumber \\
   f^{z}(\tilde{y}^{A}) & = &
   \pm R \left({\displaystyle {{1-\xi\bar\xi} \over {1+\xi\bar\xi}}}\right)
   \label{eq:xi_of_ya}
\end{eqnarray}
where the positive (negative) sign corresponds to the north (south)
patch. This defines a canonical spherical section $S_{t}$ of radius $R$
in Minkowski-space. Note that this provides a prescription for locating
the world-tube which is {\it time-independent}.

Given two 3-dimensional quantities $v^i$ and $w^i$, we introduce their
Euclidean inner product $(v\cdot w)=\delta_{ij}v^i w^j$. The
stereographic coordinates of points on the surface of the $3$-sphere
with Cartesian coordinates $x^i= (x,y,z)$ can then be
represented by $y^A(x^i)= (q,p)$, with
\begin{equation}
   q(x^i) = \frac{x}{\hat{r} \pm z} , \quad 
   p(x^i) = \pm \frac{y}{\hat{r} \pm z}
  \label{eq:stereo}
\end{equation}
on the north $(+)$ and south $(-)$ patches, where
\begin{equation}
   \hat{r}^2 = (x \cdot x).
    \label{eq:norm}
\end{equation}
Note that the stereographic coordinates are invariant under a change of
scale, $y^A(x^i)=y^A(cx^i)$. 

The Euclidean radius of the world-tube with Cartesian coordinates 
$x^{(0)a} $ is given by: 
\begin{equation}
	R^2 =(x^{(0)} \cdot x^{(0)}).
	  \label{eq:norm0}
\end{equation}
{\it This equation provides a particularly simple alternate definition of the
choice of world-tube using Cartesian coordinates. This definition
also holds for each instant of Cartesian time.}

\subsection{$4$-$d$ geometry around the world-tube}\label{sec:4dgeometry}
\label{sec:4d-geometry}

The $4$-$d$ geometry around the world-tube is fully specified by the 
$4$-$d$ metric and its derivatives (alternatively, by the metric and the
metric connection). They determine the unit normal $n^{\alpha}$ to the
$t=constant$ slices, and given the parametrization of the world-tube, the
(outward pointing) normal $s^{\alpha}$ to the world-tube. They also
determine the generator of the outgoing null radial geodesics through
the world-tube, which in turn completes the specification of the
coordinate transformation $x^{\alpha} \rightarrow \tilde{y}^{\alpha}$
in a neighborhood of the world-tube.

In practice, the necessary information is not available at the discrete
set of points on the world-tube specified by Eqs.~(\ref{eq:xi}) and
(\ref{eq:xi_of_ya}) where $\xi_{\tt ij} = \tilde{y}^{2}_{\tt i} + i
\tilde{y}^{3}_{\tt j}$ as given in Eq.~(\ref{eq:starray}).  However, the
required variables are known on the points of the computational grid
used in the simulation, a Cartesian grid $(x_{\tt i},y_{\tt j},z_{\tt k})$, from
which we interpolate them to the world-tube points to second order
accuracy.

For a standard $3+1$ simulation~\cite{york}, the variables that we need
to interpolate are the 3-d metric $g_{ij}$, the lapse $\alpha$ and the
shift $\beta^{i}$. Their spatial derivatives are also interpolated.
Their values at the world-tube points are stored for a number of time
levels, and the time derivatives of the $3$-metric, lapse and shift at
the world-tube are computed by finite-differencing between these time
levels.

Using all these values we can compute the $4$-metric $g_{\mu\nu}$, and its
first derivatives $g_{\mu\nu,\sigma}$, using the following relations:

\begin{eqnarray}
   g_{it} & = & g_{ij} \beta^{j} \quad
   \nonumber \\
   g_{tt} & = & - \alpha^{2} + g_{it} \beta^{i}
   \nonumber \\
   g_{it,\mu} & = & g_{ij,\mu} \beta^{j} + g_{ij} \beta^{j}_{,\mu}
   \nonumber \\
      g_{tt,\mu} & = & -2 \alpha \alpha_{,\mu}
      + g_{ij,\mu} \beta^{i} \beta^{j} 
      + 2 g_{ij} \beta^{i}  \beta^{j}_{,\mu} \,.
   \label{eq:ADM}
\end{eqnarray}

The unit normal $n^{\mu}$ to the hypersurface $\Sigma_{t}$ is determined
from the lapse and shift

\begin{equation}
   n^{\mu} = \frac{1}{\alpha} \left(1, -\beta^{i}\right).
\end{equation}

Let $s^{\alpha}=(s^{i},0)$ be the outward pointing unit normal to the
section $S_{t}$ of the world-tube at time $t^{n}$. By construction,
$s^{i}$ lies in the slice $\Sigma_{t}$, and it is known given the two
vectors $\partial_{\tilde{y}^{2}}$, $\partial_{\tilde{y}^{3}}$ in
$S_{t}$, defined by the parametrization of the world-tube $x^{i}(\tilde
y^{A})$

\begin{equation}
   q^{i} = {{\partial x^{i}} \over {\partial \tilde{y}^{2}}}, \quad
   p^{i} = {{\partial x^{i}} \over {\partial \tilde{y}^{3}}}. \quad
\end{equation}
These may be obtained analytically from Eq.~(\ref{eq:xi_of_ya}), the
equation for the world tube. Antisymmetrizing $q^{i}$ and $p^{i}$, we
obtain the spatial components of the normal 1-form $\sigma_{i}$ and its
norm $\sigma$
\begin{equation}
   \sigma_{i} = \epsilon_{ijk}\, q^{j} p^{k}, \quad
   \sigma = \sqrt{g^{ij} \sigma_{i} \sigma_{j}}
\end{equation}
from which $s^{i}$ is obtained by raising $\sigma_{i}$ with the
contravariant $3$-metric $g^{ij}$ on the slice $\Sigma_{t}$ and dividing by
the norm $\sigma$, yielding

\begin{equation}
   s^{i} = g^{ij} \frac{\sigma_{j}}{\sigma}.
   \label{eq:si}
\end{equation}

The generators $\ell^{\alpha}$ of the outgoing null cone $C_{t}$ through
$S_{t}$ are given on the world-tube by
\begin{equation}
   \ell^{\alpha} = {{n^{\alpha} + s^{\alpha}} 
   \over {\alpha - g_{ij} \beta^{i} s^{j}}}
   \label{eq:ell}
\end{equation}
which is normalized so that $\ell^{\alpha} t_{\alpha} = -1$, where
$t^{\alpha} = \alpha n^{\alpha} + \beta^{\alpha}$ is the Cauchy
evolution vector.

The equations in this section show explicitly how to use the output data
from the Cauchy simulation to completely reconstruct the full
$4$-geometry of the spacetime, as well as other important geometrical
objects of interest, in the neighborhood of the world-tube $\Gamma$.
This is all described in this section within the Cartesian coordinate
system used by the $3+1$ computation, and holds to the second-order
accuracy assumed for this computation. In the next three sections, we
will demonstrate how to use this information to redescribe the same
geometry in another coordinate system, the Bondi coordinates needed for
characteristic simulations. There is nothing in these sections that goes
beyond the elementary concepts of defining coordinate systems and
transforming tensors under a change of coordinates. Nevertheless,
it is quite instructive to see just how much work lies hidden beneath
the clever notation used by theorists.

\subsection{Coordinate transformation}\label{sec:coordtrans}

In this section we build the coordinate transformation between the $3+1$
Cartesian coordinates $x^{\alpha}$ and the (null) affine coordinates
$\tilde{y}^{\alpha}$. We will need this in the {\it neighborhood} of the
world-tube, not just at a {\it point} on the tube, in order to easily
pass information back and forth between the Cauchy and characteristic
computer codes in the overlap region near the world-tube. In 
Sec.~\ref{sec:nullmetric}, we will transform the metric to 
affine coordinates. Finally, in Sec.~~\ref{sec:bondimetric}, we will
complete the transformation from affine to Bondi coordinates. 

The motivation for this indirect route to the Bondi coordinate frame
deserves some comment. Both affine coordinates $\tilde{y}^{\alpha}$ and
Bondi coordinates ${y}^{\alpha}$ utilize null hypersurfaces for
foliations. Calculating these directly would require the numerical
solution of a nonlinear partial differential equation (the eikonal
equation). A much simpler, but physically equivalent, approach is to
instead solve the null geodesic equation in Cartesian coordinates, in
order to find the rays $x^{\mu}(\lambda)$ generating the required null
hypersurfaces. Secondly, we must introduce the intermediary of the
affine radial coordinate $\lambda$ rather than the Bondi surface area
coordinate $r$, since the latter is actually unknown until the
angular coordinates are defined. As shown below, it is only after the
null rays have been found that we are able to proceed with the orderly
introduction of angular coordinates $\tilde{y}^{A}$ in the exterior of
the world-tube. Finally, null geodesics can be analytically constructed
trivially in terms of $\lambda$, whereas even when we have defined the
surface area coordinate $r$, solution of the null geodesics equation
requires the numerical solution of an ordinary differential equation.
With this rationale complete, we will now proceed to carry out the
coordinate definitions and transform the metric.

By inspection, $x^{\alpha}(\lambda)$, the solution to the geodesic equation
relating $x^{\alpha}$ to $\tilde{y}^{\alpha}$ off
the world-tube is:
\begin{equation}
   x^{\alpha} = x^{(0)}{}^{\alpha} + \ell^{(0)}{}^{\alpha} \lambda
   + O(\lambda^{2}).
   \label{eq:geodesic}
\end{equation}
This expression determines $x^{\alpha}(\lambda)$ to $O(\lambda^{2})$, 
given the coefficients
\begin{equation}
   x^{(0)}{}^{\alpha} = x^{\alpha}_{\,\, |\Gamma} \quad {\rm and} \quad
   \ell^{(0)}{}^{\alpha} = x^{\alpha}_{,\lambda}{}_{\,|\Gamma} 
\end{equation}
that is, given the coordinates of the points and the
generators of the null cone through the world-tube section $S_{t}$. 
For completeness, we repeat the remaining coordinate relations
defined in Sec.~\ref{sec:world-tube}. Along each outgoing null
geodesic emerging from $S_{t}$, angular and time
coordinates are defined by setting their values to be constant along the
rays, and equal to their values on the world-tube
\begin{equation}
  \tilde{y}^{A} = y^{A}_{\,\, |\Gamma} \quad {\rm and} \quad
  \tilde{y}^{4} \equiv \tilde{u} = t. 
\end{equation}

Given the coordinate transformation
$x^{\mu}=x^{\mu}(\tilde{y}^{\alpha})$, we obtain the metric in null
affine coordinates $\tilde{\eta}_{\alpha\beta}$ by

\begin{equation}
   \tilde{\eta}_{\tilde{\alpha}\tilde{\beta}}=
   {{\partial x^{\mu}} \over {\partial \tilde{y}^{\alpha}}}
   {{\partial x^{\nu}} \over {\partial \tilde{y}^{\beta}}}
   g_{\mu\nu} .
\end{equation}
The Jacobian of the coordinate transformation is viewed as a series
expansion in the affine parameter $\lambda$ for each point on the
world-tube. Furthermore, we may omit the $\lambda$ derivatives of the
$x^{\mu}$, i.e. the $x^{\mu}_{,\lambda}$ are not needed because the
radial coordinate $\lambda$ is an affine parameter of the null
geodesics, hence the $\tilde{\eta}_{\lambda\tilde{\mu}}$ components of the null
metric are fixed:

\begin{equation}
   \tilde{\eta}_{\lambda\lambda} = \tilde{\eta}_{\lambda \tilde{A}} = 0, 
   \quad \tilde{\eta}_{\lambda \tilde{u}} = -1,
   \label{eq:knowneta}
\end{equation}
(This follows from the conditions $s^{\alpha}n_{\alpha}=0$,
$\ell^{\alpha}\ell_{\alpha}=0$, $s^{\alpha}s_{\alpha}=1$,
$n^{\alpha}n_{\alpha}=-1$ and $t^{\alpha}\ell_{\alpha}=-1$).
In other words, $\tilde{\eta}_{\tilde{\alpha}\tilde{\beta}}$ contains six independent
metric functions. The relevant part of the coordinate transformation is
then

\begin{equation}
   x^{\mu}_{,\tilde{\alpha}} \equiv
   {{\partial x^{\mu}} \over {\partial \tilde{y}^{\alpha}}} =
   x^{(0)}{}^{\mu}_{,\tilde{\alpha}} + x^{(1)}{}^{\mu}_{,\tilde{\alpha}} \lambda
   + O(\lambda^{2}), 
   \quad x^{(1)}{}^{\mu}_{,\tilde{\alpha}} \equiv \ell^{(0)}{}^{\mu}_{,\tilde{\alpha}} ,
   \quad {\rm for} \quad \tilde{\alpha} = (\tilde{A},\tilde{u}).
	\label{eq:jacob}
\end{equation}

Because the specification of the world-tube is time-independent, and
the slices $S_{t}$ of the world-tube are by construction at $t=constant$,
only the angular derivatives of the $x^{(0)}{}^{i}$ for $i=1,2,3$
survive, i.e. the $O(\lambda^{0})$ part of the Jacobian is given by the
condition $\partial t/\partial \tilde{u}_{|\Gamma} = 1$, and by the relations

\begin{equation}
   x^{(0)}{}^{i}_{,\tilde{A}} = {{\partial f^{i}(\tilde{y}^{B})} \over {\partial \tilde{y}^{A}}},
\end{equation}
which can be evaluated from the analytic
expressions Eq.~(\ref{eq:xi_of_ya}).

To evaluate the $O(\lambda)$ part of the Jacobian, we note that

\begin{equation}
   x^{\mu}_{,\lambda \tilde{A}} = \ell^{\mu}_{,\tilde{A}}, \quad
   x^{\mu}_{,\lambda \tilde{u}} = \ell^{\mu}_{,\tilde{u}} 
   \label{eq:j1}
\end{equation}

We will spend the remainder of this section evaluating these terms. To
proceed, we see from Eq.~(\ref{eq:ell}) that this will require
derivatives of $n^{\mu}$ and $s^{i}$. For the simple case of geodesic
slicing, $\alpha=1$, $\beta^{i}=0$, the derivatives of the normal
$n^{\mu}$ vanish, and the transformation is contained purely in the
derivatives of $s^{\mu}$. In general, the angular derivatives of
$n^{\mu}$ at $\lambda=constant$ can be computed in terms of spatial 3+1
derivatives of the lapse and shift, transformed with the
$O(\lambda^{0})$ Jacobian, i.e.

\begin{equation}
   n^{\mu}_{,\tilde{A}} = n^{\mu}_{,j} x^{j}_{,\tilde{A}}
   \label{eq:naA}
\end{equation}
The 3+1 derivatives are given by

\begin{eqnarray}
   n^{i}_{,j} &=& {1 \over \alpha^{2}} 
   \left( \alpha_{,j} \beta^{i} - \alpha \beta^{i}_{,j} \right)
   \nonumber \\
   n^{t}_{,j} &=& - {1 \over \alpha^{2}} \alpha_{,j} .
   \label{eq:naj}
\end{eqnarray}
The retarded time derivative $\partial_{\tilde{u}}$ at $\lambda=constant$ is
simply the $3+1$ time derivative $\partial_{t}$, therefore
$n^{\mu}_{,\tilde{u}} = n^{\mu}_{,t}$ where

\begin{eqnarray}
   n^{i}_{,t} &=& {1 \over \alpha^{2}} 
   \left( \alpha_{,t} \beta^{i} - \alpha \beta^{i}_{,t} \right) ,
   \nonumber \\
   n^{t}_{,t} &=& - {1 \over \alpha^{2}} \alpha_{,t} .
   \label{eq:nat}
\end{eqnarray}

From Eq.~(\ref{eq:si}), and since the $\sigma_{k}$ are time-independent,
the time derivative of $s^{i}$ is given by

\begin{eqnarray}
   s^{i}_{,t} &=& g^{ik}_{,t} {\sigma_{k} \over \sigma}
              - g^{ik} {{\sigma_{k} \sigma_{,t}} \over \sigma^{2}} 
              = -g^{im} g^{kn} g_{mn,t} {\sigma_{k} \over \sigma}
              - s^{i} {\sigma_{,t} \over \sigma}  \nonumber \\
              &=& \mbox{} -g^{im} g_{mn,t} s^{n} 
              - s^{i} {\sigma_{,t} \over \sigma}
\end{eqnarray}
where the time derivative of $\sigma$ can be calculated from

\begin{equation}
   2 \sigma \sigma_{,t} = \left( \sigma^2 \right)_{,t} =
   g^{kl}_{,t} \sigma_{k} \sigma_{l} =
   - g^{km} g^{ln} g_{mn,t} \sigma_{k} \sigma_{l} =
   - s^{m} s^{n} g_{mn,t} \sigma^{2} ,
   \label{eq:dsigmadt}
\end{equation}
with the resulting expression

\begin{equation}
   s^{i}_{,t} = \left( -g^{im} + s^{i} \frac{1}{2} s^{m} \right) 
   g_{mn,t} s^{n} .
   \label{eq:sit}
\end{equation}

Similarly, it follows from Eq.~(\ref{eq:si}) that 

\begin{eqnarray}
   s^{i}_{,\tilde{A}} &=& g^{ik}_{,j} x^{j}_{,\tilde{A}} {\sigma_{k} \over \sigma}
   + g^{ik} {\sigma_{k,\tilde{A}} \over \sigma}
   - g^{ik} {\sigma_{k} \sigma_{,\tilde{A}} \over \sigma^{2}} \nonumber \\
   &=& \mbox{} -g^{in} g^{km} g_{mn,j} \,x^{j}_{,\tilde{A}} {\sigma_{k} \over \sigma}
   + g^{ik} {\sigma_{k,\tilde{A}} \over \sigma} 
   - s^{i} {\sigma_{,\tilde{A}} \over \sigma}
   \label{eq:siA1}
\end{eqnarray}
where the $\sigma_{k,\tilde{A}}$ are obtained from the analytic
expressions Eq.~(\ref{eq:xi_of_ya}), and $\sigma_{,\tilde{A}}$ from

\begin{eqnarray}
   2 \sigma \sigma_{,\tilde{A}} & = & \left( \sigma^2 \right)_{,\tilde{A}}
   = \left( g^{kl} \sigma_{k} \sigma_{l} \right)_{,\tilde{A}}
   = g^{kl}_{,j} x^{j}_{,\tilde{A}} \sigma_{k} \sigma_{l}
   + 2 \, g^{kl} \sigma_{l} \sigma_{k,\tilde{A}} \nonumber \\
   &=& \mbox{} - g^{km} g^{ln} g_{mn,j} x^{j}_{,} \sigma_{k} \sigma_{l}
   + 2 \, g^{kl} \sigma_{l} \sigma_{k,\tilde{A}}
   \nonumber \\
   & = & \mbox{} - s^{m} s^{n} g_{mn,j} x^{j}_{,\tilde{A}} \sigma^{2}
   + 2 s^{k} \sigma \sigma_{k,\tilde{A}}
   \label{eq:sigmaA}
\end{eqnarray}

Collecting Eqs.~(\ref{eq:siA1}) and (\ref{eq:sigmaA}), we arrive at the
angular derivatives of the normal to the world-tube

\begin{eqnarray}
   s^{i}_{,\tilde{A}} & = & -g^{in} s^{m} g_{mn,j} \,x^{j}_{,\tilde{A}} 
   + g^{ik} {\sigma_{k,\tilde{A}} \over \sigma}
   + s^{i} \left( \frac{1}{2} s^{m} s^{n} g_{mn,j} x^{j}_{,\tilde{A}} 
                  - s^{k} {\sigma_{k,\tilde{A}} \over \sigma} \right)
   \nonumber \\
   & = & \left( g^{in} - s^{i} s^{n} \right) {\sigma_{n,\tilde{A}} \over \sigma}
   + \left( - g^{in} + \frac{1}{2} s^{i} s^{n} \right) 
     s^{m} g_{mn,j} x^{j}_{,}
   \label{eq:siA}
\end{eqnarray}

\subsection{Null metric \protect{$\tilde{\eta}_{\alpha\beta}$}}
\label{sec:nullmetric}

Given the $4$-metric and its Cartesian derivatives at the world-tube, we
can calculate its derivative with respect to the affine parameter
$\lambda$ according to

\begin{equation}
   g_{\alpha\beta,\lambda}{}_{|\Gamma} = g^{(0)}_{\alpha\beta,\mu} 
  \ell^{(0)}{}^{\mu}
   \label{eq:g1}
\end{equation}

Having obtained the relevant parts of the coordinate transformation
$\tilde{y}^{\alpha} \rightarrow x^{\alpha}$, Eqns. (\ref{eq:j1}),
(\ref{eq:naA}), (\ref{eq:naj}), (\ref{eq:nat}), (\ref{eq:sit}), and
(\ref{eq:siA}), and given the metric and its $\lambda$ derivative as
per Eq. (\ref{eq:g1}), we can expand the null metric as follows

\begin{equation}
   \tilde{\eta}_{\tilde{\alpha}\tilde{\beta}} = \tilde{\eta}^{(0)}_{\tilde{\alpha}\tilde{\beta}}
   + \tilde{\eta}_{\tilde{\alpha}\tilde{\beta},\lambda} \lambda + O(\lambda^{2}),
\end{equation}
where the coefficients are given by

\begin{eqnarray}
   \tilde{\eta}^{(0)}_{\tilde{u}\tilde{u}} & = & g_{tt}{}_{|\Gamma}
   \nonumber \\
   \tilde{\eta}^{(0)}_{\tilde{u}\tilde{A}} & = & x^{i}_{,\tilde{A}} g_{it}{}_{|\Gamma} 
   \nonumber \\
   \tilde{\eta}^{(0)}_{\tilde{A}\tilde{B}} & = & x^{i}_{,\tilde{A}} x^{j}_{,\tilde{B}} g_{ij}{}_{|\Gamma} 
   \label{eq:eta0}
\end{eqnarray}
and, for the $\lambda$ derivative

\begin{eqnarray}
   \tilde{\eta}_{\tilde{u}\tilde{u},\lambda} & = & \left[g_{tt,\lambda} 
   + 2\, \ell^{\mu}_{,\tilde{u}} g_{\mu t}\right]_{|\Gamma} + O(\lambda)
   \nonumber \\
   \tilde{\eta}_{\tilde{u} \tilde{A},\lambda} & = & \left[ x^{k}_{,\tilde{A}} \left(
      \ell^{\mu}_{,\tilde{u}} g_{k\mu} + g_{kt,\lambda} \right)
      + \ell^{k}_{,\tilde{A}} g_{kt} + \ell^{t}_{,\tilde{A}} g_{tt} \right]_{|\Gamma} 
      + O(\lambda) , 
   \nonumber \\
   \tilde{\eta}_{\tilde{A}\tilde{B},\lambda} & = & \left[ x^{k}_{,\tilde{A}} x^{l}_{,\tilde{B}} g_{kl,\lambda}
   + \left(  \ell^{\mu}_{,\tilde{A}} x^{l}_{,\tilde{B}} + \ell^{\mu}_{,\tilde{B}} x^{l}_{,\tilde{A}} \right) 
     g_{\mu l} \right]_{|\Gamma}
   + O(\lambda) .
   \label{eq:eta1}
\end{eqnarray}
The remaining components are fixed by Eq.~(\ref{eq:knowneta}).

It is worthwhile to discuss a subtlety in the rationale underlying the
computational strategy used here. The purpose of carrying out an
expansion in $\lambda$ is to enable us to give the metric variables in a
small region off the world-tube, {\it i.e.} at points of the grid used
to discretize the null equations. This method is an alternative to the
more obvious and straightforward strategy of interpolation to determine
metric values at needed points on the null computational grid. At first
glance, it might appear a more cumbersome way to proceed, due to the
necessity of analytic calculation of the derivative of the metric on the
world tube to determine coefficients needed for the metric expansion in
powers of $\lambda$. On the contrary, however, {\it this trick allows a
major simplification in computer implementation.} It allows us to use a
null code which does not require any special implementation at the
irregular boundary defined by the world-tube, and it automatically
ensures continuity of the metric and the extrinsic curvature at the
world-tube. The obvious alternative of a brute-force interpolation
approach would require a full $4$-dimensional evaluation with quite
complicated logic, since it would lie at the edge of both the Cauchy and
null computational grids. Instead, the $\lambda$ expansion reduces the
complexity by one dimension, allowing for much easier numerical
implementation.

We also need to compute the contravariant null metric,
$\tilde{\eta}{}^{\alpha\beta}$, which we similarly consider as an expansion
in powers of $\lambda$,

\begin{equation}
   \tilde{\eta}^{\tilde{\mu}\tilde{\nu}} = \tilde{\eta}^{(0)}{}^{\tilde{\mu}\tilde{\nu}}
   + \tilde{\eta}^{\tilde{\mu}\tilde{\nu}}_{,\lambda} \lambda
   + O(\lambda^{2}) ,
\end{equation}
with coefficients given by

\begin{equation}
   \tilde{\eta}^{(0)}{}^{\tilde{\mu}\tilde{\alpha}} 
   \tilde{\eta}^{(0)}_{\tilde{\alpha}\tilde{\nu}} = \delta^{\tilde{\mu}}_{\tilde{\nu}} , \quad
   \tilde{\eta}^{\tilde{\mu}\tilde{\nu}}_{,\lambda}  = - \tilde{\eta}^{\tilde{\mu}\tilde{\alpha}} \,
   \tilde{\eta}^{\tilde{\beta}\tilde{\nu}} \, \tilde{\eta}_{\tilde{\alpha}\tilde{\beta},\lambda} .
\end{equation}

It follows from Eq.~(\ref{eq:knowneta}) that the following components
of the contravariant null metric in the $\tilde{y}^{\alpha}$ coordinates are
fixed

\begin{equation}
   \tilde{\eta}^{\lambda \tilde{u}} = -1 \quad
   \tilde{\eta}^{\tilde{u}\tilde{A}} = \tilde{\eta}^{\tilde{u}\tilde{u}} = 0, 
   \label{eq:tildeup}
\end{equation}
therefore the contravariant null metric can be computed by

\begin{eqnarray}
   \tilde{\eta}^{\tilde{A}\tilde{B}} \tilde{\eta}_{\tilde{B}\tilde{C}} & = & \delta^{\tilde{A}}_{\, \,\tilde{C}}, 
   \nonumber \\
   \tilde{\eta}^{\lambda \tilde{A}} & = & \tilde{\eta}^{\tilde{A}\tilde{B}} \tilde{\eta}_{\tilde{B}\tilde{u}}
   \nonumber \\
   \tilde{\eta}^{\lambda\lambda} & = & - \tilde{\eta}_{\tilde{u}\tilde{u}}
   + \tilde{\eta}^{\lambda \tilde{A}} \tilde{\eta}_{\tilde{A}\tilde{u}}
   \label{eq:etaup0}
\end{eqnarray}
and similarly for its $\lambda$ derivative

\begin{eqnarray}
   \tilde{\eta}^{\tilde{A}\tilde{B}}_{,\lambda} & = & - \tilde{\eta}^{\tilde{A}\tilde{C}} \tilde{\eta}^{\tilde{B}\tilde{D}} 
   \tilde{\eta}_{\tilde{C}\tilde{D},\lambda}
   \nonumber \\
   \tilde{\eta}^{\lambda \tilde{A}}_{,\lambda} & = & \tilde{\eta}^{\tilde{A}\tilde{B}}
   \left( \tilde{\eta}_{\tilde{u}\tilde{B},\lambda} - \tilde{\eta}^{\lambda \tilde{C}} 
   \tilde{\eta}_{\tilde{C}\tilde{B},\lambda} \right)
   \nonumber \\
   \tilde{\eta}^{\lambda\lambda}_{,\lambda} & = & - \tilde{\eta}_{\tilde{u}\tilde{u},\lambda} 
   + 2\, \tilde{\eta}^{\lambda \tilde{A}} \tilde{\eta}_{\tilde{u}\tilde{A},\lambda} 
   - \tilde{\eta}^{\lambda \tilde{A}} \tilde{\eta}^{\lambda \tilde{B}} 
   \tilde{\eta}_{\tilde{A}\tilde{B},\lambda} 
\end{eqnarray}

\subsection{Metric in Bondi coordinates}\label{sec:bondimetric}

The surface area coordinate
$r(u,\lambda,\tilde{y}^{A})$ is defined by

\begin{equation}
   \label{eq:rofeta}
   r  =  \left( \frac{det(\tilde{\eta}_{\tilde{A}\tilde{B}})} {det(q_{AB})} \right)
   ^{\frac{1}{4}},
   \label{eq:r}
\end{equation}
where, for our choice of stereographic
coordinates $\xi= q + ip $, we use $\tilde{y}^{A} = (q,p)$ and 
$\det(q_{AB})=16/(1+q^{2}+p^{2})^{4}$ .  To carry out the coordinate transformation
$\tilde{y}^{\alpha} \rightarrow y^{\alpha}$ on the null metric, where
$y^{\alpha}=(r,y^{A},u)$, we need to know $r_{,\lambda}$, $r_{,\tilde{A}}$ and
$r_{,\tilde{u}}$. From Eq.~(\ref{eq:rofeta}) it follows

\begin{equation}
   r_{,\lambda} = \frac{r}{4} \tilde{\eta}^{\tilde{A}\tilde{B}} \tilde{\eta}_{\tilde{A}\tilde{B},\lambda} .
   \label{eq:rl}
\end{equation}
Similarly, 

\begin{equation}
   r_{,\tilde{C}} = \frac{r}{4} \left(\tilde{\eta}^{\tilde{A}\tilde{B}} \tilde{\eta}_{\tilde{A}\tilde{B},\tilde{C}} 
   - \frac{det(q_{\tilde{A}\tilde{B}})_{,\tilde{C}}}{det(q_{\tilde{A}\tilde{B}})} \right),
\end{equation}
where
\begin{eqnarray}
   \frac{det(q_{\tilde{A}\tilde{B}})_{,\tilde{C}}}{det(q_{\tilde{A}\tilde{B}})} & = & 
   - {\displaystyle{8 \over {1 + q^{2} + p^{2}}}} \tilde{y}^{\tilde{C}}
   \nonumber \\
   \tilde{\eta}_{\tilde{A}\tilde{B},\tilde{C}} & = & 
   \left( x^{i}_{,\tilde{A}\tilde{C}}\, x^{j}_{,\tilde{B}}
        + x^{i}_{,\tilde{A}}\,  x^{j}_{,\tilde{B}\tilde{C}} \right) g_{ij}
   + x^{i}_{,\tilde{A}}\, x^{j}_{,\tilde{B}}\, x^{k}_{,\tilde{C}}\, g_{ij,k}
\end{eqnarray}
with the $x^{i}_{,\tilde{A}\tilde{C}}$ given functions of $(q,p)$. From Eqs.~(\ref{eq:r}) and 
(\ref{eq:eta0})

\begin{equation}
   r_{,\tilde{u}} =  \frac{r}{4}\, \tilde{\eta}^{\tilde{A}\tilde{B}} \tilde{\eta}_{\tilde{A}\tilde{B},\tilde{u}}
\end{equation}
where
\begin{equation}
   \tilde{\eta}_{\tilde{A}\tilde{B},\tilde{u}} = \left[ x^{i}_{,\tilde{A}}\, x^{j}_{,\tilde{B}}\, g_{ij,t} \right]
   _{|\Gamma} + O(\lambda) .
\end{equation}

The null metric $\eta^{\alpha\beta}$ in Bondi coordinates is defined 
on the world-tube $\Gamma$ by

\begin{equation}
   \eta^{\alpha\beta} = 
   {{\partial y^{\alpha}} \over {\partial \tilde{y}^{\mu}}}
   {{\partial y^{\beta}} \over {\partial \tilde{y}^{\nu}}}
   \tilde{\eta}^{\tilde{\mu}\tilde{\nu}}
   \label{eq:etaup}
\end{equation}
Note that the metric of the sphere is unchanged by this coordinate
transformation, i.e.  $\eta^{AB} = \tilde{\eta}^{\tilde{A}\tilde{B}} $, so we need to
compute only the elements $\eta^{rr}$, $\eta^{rA}$ and $\eta^{ru}$ on
$\Gamma$, or
equivalently the metric functions $\beta$, $U^{A}$ and $V$.  From
Eq.~(\ref{eq:tildeup}),

\begin{eqnarray}
   \eta^{rr} & = & r_{,\tilde{\alpha}}\, r_{,\tilde{\beta}}\, \tilde{\eta}^{\tilde{\alpha}\tilde{\beta}} 
   = \left(r_{,\lambda}\right)^2 \tilde{\eta}^{\lambda \lambda} 
   + 2\, r_{,\lambda}\, \left(r_{,\tilde{A}}\,\tilde{\eta}^{\lambda \tilde{A}} - r_{,\tilde{u}} \right)
   + r_{,\tilde{A}}\,r_{,\tilde{B}}\, \tilde{\eta}^{\tilde{A}\tilde{B}}
   \nonumber \\
   \eta^{rA} & = & r_{,\tilde{\alpha}}\, \tilde{\eta}^{\tilde{\alpha} \tilde{A}} 
   = r_{,\lambda}\, \tilde{\eta}^{\lambda \tilde{A}} + r_{,\tilde{B}}\, \tilde{\eta}^{\tilde{A}\tilde{B}}
   \nonumber \\
   \eta^{ru} & = & r_{,\tilde{\alpha}}\, \tilde{\eta}^{\tilde{\alpha} \tilde{u}} = - r_{,\lambda}
   \label{eq:bondim}
\end{eqnarray}

The contravariant Bondi metric can be written in the form
\begin{equation}
\eta^{\alpha\,\beta} =
\left [
\begin {array}{cccc} 
 e^{-2\,\beta} \displaystyle{\frac{V}{r}} &
-e^{-2\,\beta} U^{2} &
-e^{-2\,\beta} U^{3} &
-e^{-2\,\beta}
\\
\noalign{\medskip}
-e^{-2\,\beta} U^{2} &
 r^{-2}\, h^{22} &
 r^{-2}\, h^{23} &
 0
\\
\noalign{\medskip}
-e^{-2\,\beta} U^{3} &
 r^{-2}\, h^{32} &
 r^{-2}\, h^{33} &
 0
\\
\noalign{\medskip}
-{e^{-2\,\beta}} &
 0 &
 0 &
 0
\end {array}
\right] ,
\label{eq:bondimetric}
\end{equation}
where $h_{AB}$ is a metric on the sphere of surface area $4\pi$, such
that $h_{AB}h^{BC}=\delta_{A}^{\, \, C}$ and $det(h_{AB})=det(q_{AB})=q$, for
$q_{AB}$ a unit sphere metric.

\subsection{Bondi Variables for starting up the Null code at the world-tube}
\label{sec:nullbdry}

The natural variables of the null formalism are certain combinations of
the null metric functions, which we will give in this section.  They
will be expressed as an expansion in $\lambda$, as was done in the
previous sections, to enable us to give these variables in a small
region off the world-tube, {\it i.e.} at points of the grid used to
discretize the null equations. {\it These expressions are the main results
of the extraction module.}

\subsubsection{The metric of the sphere $J$}

Given $r$ and $r_{,\lambda}$, and noting that
\begin{equation}
\eta_{AB} = \tilde{\eta}_{AB} \equiv r^2 h_{AB},
\end{equation}

\begin{eqnarray}
   h_{AB} &=& \frac{1}{r^{2}} \eta_{AB}
   \nonumber \\
   h_{AB,\lambda} &=& \frac{1}{r^{2}}
   \left( \eta_{AB,\lambda} 
   - \frac{2\,r_{,\lambda}}{r}\,\eta_{AB}\right)
\end{eqnarray}

In terms of $q^{A}$, $\bar{q}^{A}$ and the metric on the unit sphere
$h_{AB}$, we define the metric functions

\begin{equation}
   J \equiv \frac{1}{2} q^{A} q^{B} h_{AB}, \quad
   K \equiv \frac{1}{2} q^{A} \bar q^{B} h_{AB}, \quad
   K^2 = 1 + J \bar J .
   \label{eq:jdef}
\end{equation}
It suffices to evaluate $J$, since the last relation holds for a Bondi
metric.  We give expressions for $J$ and $J_{,\lambda}$

\begin{eqnarray}
   J & = & \frac{1}{2\,r^{2}} q^{A} q^{B} \eta_{AB} 
   \nonumber \\ 
   J_{,\lambda} & = & \frac{1}{2\,r^{2}} q^{A} q^{B} \eta_{AB,\lambda}
   - 2\,\frac{r_{,\lambda}}{r} J
   \label{eq:jl}
\end{eqnarray}

Then, in the neighborhood of $\Gamma$, the metric of the sphere is
explicitly given in Bondi coordinates  to second-order
accuracy by these last two expressions as
\begin{equation}
	J(y^\alpha) = J + J_{,\lambda} \lambda + O(\lambda^{2}).
\end{equation}

\subsubsection{The ``expansion factor'' $\beta$}

From the last of Eq.~(\ref{eq:bondim}) we obtain the metric function $\beta$

\begin{equation}
   \beta = -\frac{1}{2}\log(r_{,\lambda}). 
\end{equation}
We want to know also its $\lambda$-derivative, but instead of
calculating $\eta^{\lambda u}_{,\lambda}$ directly (which would involve
$r_{,\lambda\lambda}$)

\begin{equation}
   \beta_{,\lambda} = - \frac{\eta^{ru}_{,\lambda}}{2\,\eta^{ru}}
   =  - \frac{r_{,\lambda\lambda}}{2\,r_{,\lambda}}
   \label{eq:rlls}
\end{equation}
we obtain $\beta_{,\lambda}$ from the characteristic equation

\begin{equation}
   \beta_{,r} = \frac{r}{8} \left( J_{,r} \bar{J}_{,r} 
   - \left(K_{,r}\right)^{2} \right)
   \label{eq:betar}
\end{equation}
At constant angles $(q,p)$, the relation $\partial_{\lambda}
= r_{,\lambda} \partial_{r}$ holds, and we know from Eq.~(\ref{eq:jl}) $J$
and $J_{,\lambda}$ for each outgoing radial null geodesic through the
world-tube, thus we can write

\begin{equation}
   \beta_{,\lambda} = \frac{r}{8\, r_{,\lambda}} 
   \left( J_{,\lambda} \bar{J}_{,\lambda} - \left(K_{,\lambda}\right)^{2} 
   \right)
   \label{eq:betal}
\end{equation}
and from Eq.~(\ref{eq:jdef}), it follows that

\begin{equation}
   K_{,\lambda} = \frac{1}{K} \Re \left(\bar{J} J_{,\lambda} \right) .
   \label{eq:Kl}
\end{equation}

\begin{equation}
   \beta_{,\lambda} = \frac{r}{8\, r_{,\lambda}} 
   \left( J_{,\lambda} \bar{J}_{,\lambda} - 
   \frac{1}{1 + J \bar{J}} 
   \left[\Re \left(\bar{J} J_{,\lambda} \right)\right]^2  \right) .
   \label{eq:nbetal}
\end{equation}
Then, $\beta$ is found to second-order accuracy by:
\begin{equation}
	\beta(y^\alpha) = \beta + \beta_{,\lambda} \lambda + O(\lambda^{2}).
\end{equation}

\subsubsection{The ``shift'' $U$}

The metric function $U$ is related to the Bondi metric, 
Eq.~(\ref{eq:bondimetric}) by
\begin{equation}
   U \equiv U^{A} q_{A} = \frac{\eta^{rA}}{{\eta^{ru}}} q_{A}
     = - \left(\tilde{\eta}^{\lambda \tilde{A}} 
       + \frac{r_{,\tilde{B}}}{r_{,\lambda}} \tilde{\eta}^{\tilde{A}\tilde{B}} \right) q_{\tilde{A}} ,
\end{equation}
where we have made use of Eq.~(\ref{eq:bondim}).
We also want the $\lambda$ derivative of $U$

\begin{eqnarray}
   U_{,\lambda} & = & - \left[ \tilde{\eta}^{\lambda \tilde{A}}_{,\lambda}
       + \left( \frac{r_{,\lambda \tilde{B}}}{r_{,\lambda}} 
              - \frac{r_{,\tilde{B}}\,r_{,\lambda\lambda}}{r^2_{,\lambda}}
         \right) \tilde{\eta}^{\tilde{A}\tilde{B}}
       + \frac{r_{,\tilde{B}}}{r_{,\lambda}} \tilde{\eta}^{\tilde{A}\tilde{B}}_{,\lambda} 
                    \right] q_{\tilde{A}} ,
   \nonumber \\
   & = & - \left( \tilde{\eta}^{\lambda \tilde{A}}_{,\lambda}
                 + \frac{r_{,\lambda \tilde{B}}}{r_{,\lambda}} \tilde{\eta}^{\tilde{A}\tilde{B}}
                 + \frac{r_{,\tilde{B}}}{r_{,\lambda}} \tilde{\eta}^{\tilde{A}\tilde{B}}_{,\lambda} 
                    \right) q_{\tilde{A}}
         + 2 \, \beta_{,\lambda} \left( U 
                     + \tilde{\eta}^{\lambda \tilde{A}} q_{\tilde{A}} \right) ,
\end{eqnarray}
where in the last line we have used Eq.~(\ref{eq:rlls}) to eliminate
$r_{,\lambda\lambda}$.

Then, $U$ is found to second-order accuracy by:
\begin{equation}
	U(y^\alpha) = U + U_{,\lambda} \lambda + O(\lambda^{2}).
\end{equation}

\subsubsection{The ``mass aspect'' $W$}

The metric function $V$ is given in terms of $\eta^{rr}$ and $\eta^{ru}$ by 
\begin{equation}
   V \equiv -r\, \eta^{rr}/\eta^{ru} . 
\end{equation}
For Minkowski-space, $V=r$ at infinity. Thus, we introduce the auxiliary
metric function $W \equiv (V-r)/r^2$, which is regular at null infinity. In
terms of the contravariant null metric (with the affine parameter $\lambda$
as the radial coordinate)

\begin{equation}
   W = \frac{1}{r} \left(\frac{\eta^{rr}}{r_{,\lambda}} - 1 \right)
     = \frac{1}{r} \left( r_{,\lambda} \tilde{\eta}^{\lambda\lambda}
   + 2 \, \left( r_{,\tilde{A}}\,\tilde{\eta}^{\lambda \tilde{A}} - r_{,u} \right)
   + \frac{r_{,\tilde{A}}\,r_{,\tilde{B}}}{r_{,\lambda}}\, \tilde{\eta}^{\tilde{A}\tilde{B}} - 1 \right)
\end{equation}

The $\lambda$ derivative of $W$ is given by

\begin{eqnarray}
   W_{,\lambda} & = & - \frac{r_{,\lambda}}{r} W
   + \frac{1}{r} \left( r_{,\lambda} \tilde{\eta}^{\lambda\lambda}
   + 2 \, \left( r_{,\tilde{A}}\,\tilde{\eta}^{\lambda \tilde{A}} - r_{,u} \right)
   + \frac{r_{,\tilde{A}}\,r_{,\tilde{B}}}{r_{,\lambda}}\, \tilde{\eta}^{\tilde{A}\tilde{B}} - 1 
   \right)_{,\lambda} \nonumber
   \\
   & = &  - \frac{r_{,\lambda}}{r} 
   \left(
         \left( \frac{r_{,\lambda}}{r} + 2\, \beta_{,\lambda} \right)
         \tilde{\eta}^{\lambda\lambda}
         - \tilde{\eta}^{\lambda\lambda}_{,\lambda} 
         - \frac{1}{r}
   \right)
   + \frac{2}{r} 
     \left( \frac{r_{,\lambda}r_{,u}}{r} - r_{,\lambda u} \right)
   \nonumber \\
   &  & + \frac{2}{r} 
         \left( 
               r_{,\lambda \tilde{A}}
               - \frac{r_{,\lambda}r_{,\tilde{A}}}{r} 
         \right)
         \tilde{\eta}^{\lambda \tilde{A}}
   + 2\frac{r_{,\tilde{A}}}{r}\,\tilde{\eta}^{\lambda \tilde{A}}_{,\lambda} 
   \nonumber \\
   &  & +  \frac{r_{,\tilde{B}}}{r\,r_{,\lambda}} 
     \left(
           2 \, r_{,\lambda \tilde{A}} \tilde{\eta}^{\tilde{A}\tilde{B}}
         + 2 \, \beta_{,\lambda} r_{,\tilde{A} }
         + r_{,\tilde{A}} \tilde{\eta}^{\tilde{A} \tilde{B}}_{,\lambda} 
     \right)
   - \frac{r_{,\tilde{A}} \, r_{,\tilde{B}}}{r^2}\, \tilde{\eta}^{\tilde{A}\tilde{B}} 
\end{eqnarray}

Then, $W$ is found to second-order accuracy by:
\begin{equation}
	W(y^\alpha) = W + W_{,\lambda} \lambda + O(\lambda^{2}).
\end{equation}

\section{Injection}

We have now reached the halfway point in the problem of relating data
between the Null and Cauchy computational approaches. We turn to
injection, the final part of the problem. Injection is the inverse
problem to extraction, and allows us to use data obtained by
characteristic evolution in the exterior of the world-tube in order to
provide the exact boundary conditions for Cauchy evolution in its interior.
That is, given the metric in Bondi coordinates in a region around the
world-tube $\Gamma$, two steps must be carried out. The first ingredient
is to relate the Cartesian coordinate system to the Bondi
null coordinates. Secondly, the Cartesian metric components must be
calculated in the neighborhood of $\Gamma$ in the usual way, from the
Bondi metric components and the Jacobian of the coordinate
transformation. 

\subsection{Locating the Cartesian grid worldlines}

Given a worldline of a Cartesian grid point, we need to locate, in null
coordinates, its intersection with a given null hypersurface.
Mathematically, this means: Given $x^i$ and $u$ we want to find $t$ and
the null coordinates $r$, as well as $y^A=y^A(x^{(0)i})=(q,p)$
(where, at this stage, $x^{(0)i}$ is unknown). 

The starting point is the geodesic equation Eq.~(\ref{eq:geodesic})
\begin{equation}
  x^\alpha =x^{(0)\alpha}+\lambda \ell^{(0)\alpha}+O(\lambda^2),
\label{eq:geo}
\end{equation} 
Here $\ell^{(0)\alpha}=\ell^{\alpha}(u,y^A(x^{(0)i})$ is the value
of the null vector given by Eq.~(\ref{eq:ell}) at the point on the
world-tube with null coordinates $u=t^{(0)}$, $\lambda =0$  and
$y^A(x^{(0)i})$.

We set $L^\alpha \equiv \ell^{\alpha}(u,y^A(x^i)),$
so that 
\begin{equation}
  \ell^{(0)\alpha} \equiv \ell^{\alpha}(x^{(0)\beta}))  = L^\alpha+O(\lambda), 
\end{equation}
since $y^A(x^i)=y^A(x^{(0)i})+O(\lambda)$. ($L^\alpha$ can now be
interpolated onto the stereographic coordinate patch to fourth order
accuracy in terms of known quantities near the world tube.)

To the same approximation as (\ref{eq:geo}), we have
\begin{equation}
   x^\alpha=x^{(0)\alpha}+\lambda L^\alpha+O(\lambda^2)
\end{equation}  
so that we have four equations for the five unknowns $(x^{(0)i}, t,$ and $\lambda)$:
\begin{eqnarray}
   t & = & u+\lambda L^t+O(\lambda^2) \\
\label{eq:Lgeot}
   x^i & = & x^{(0)i}+\lambda L^i+O(\lambda^2).
\label{eq:Lgeoi}
\end{eqnarray}

To find the fifth unknown, we need one additional equation introducing
new information. This is conveniently supplied by the equation defining
the world-tube Eq.~(\ref{eq:norm0}). Even though we have not yet found
the values for $x^{(0)i}$, we may use Eq.~(\ref{eq:norm0}) to eliminate
these unknowns by taking the Euclidean norm of Eq.~(\ref{eq:Lgeoi}), whence
using Eq.~(\ref{eq:norm}) we find:

\begin{equation}
  R^{2} = \hat{r}^2 +\lambda^{2} (L \cdot L) 
   - 2 \lambda (\L \cdot x) +O(\lambda^2).
\label{eq:lambda}
\end{equation}
where the value of $R^2=(x^{(0)} \cdot x^{(0)})$ is known from the beginning by the
definition of the world tube, and $\hat{r}^2 =  (x \cdot x).$

This is a quadratic in $\lambda$, but we are consistently only keeping
terms to linear order. This leads to 
\begin{equation}
\Lambda = (\hat{r}^{2} - R^{2})/ \left[ 2 (L \cdot x) \right] 
\end{equation}

Consequently, $\Lambda=\lambda+O(\lambda^2)$ and can be calculated in
terms of known quantities. This leaves four unknown quantities
remaining. To proceed to evaluate these in terms of known data, we set
$T=u+\Lambda L^t, X^{(0)i}=(X^{(0)},Y^{(0)},Z^{(0)}) =x^i-\Lambda L^i$,
and $Y^A= (Q,P) = y^A(X^{(0)i})$. Then, using Eq.~(\ref{eq:stereo})
\begin{equation}
  Q= \frac{X^{(0)}}{R \pm Z^{(0)}}, \quad P = \pm \frac{Y^{(0)}}{R \pm Z^{(0)}},
\end{equation} 
on the north $(+)$ and south $(-)$ patches. $T$, $\Lambda$ and $Y^A$ are accurate to $O(\lambda^2)$ since
$t=T+O(\lambda^2)$, $\lambda=\Lambda+O(\lambda^2)$ and
$y^A=Y^A+O(\lambda^2)$.

Finally, the value of $r$ is obtained from
\begin{equation}
   r =  r^{(0)} + r^{(0)}_{,\lambda} \Lambda , \quad.
\end{equation}
and the values of $r_{,\lambda}$, $r_{,A}$ and $r_{,u}$ (used in
computing the $\tilde{\eta}^{\tilde{\mu}\tilde{\nu}}$ metric) from
\begin{equation}
   r_{,\lambda} =  r^{(0)}_{,\lambda} + r^{(0)}_{,\lambda\lambda} \Lambda, \quad
   r_{,A} =  r^{(0)}_{,A} + r^{(0)}_{,\lambda A} \Lambda, \quad
   r_{,u} =  r^{(0)}_{,u} + r^{(0)}_{,\lambda u} \Lambda 
\end{equation}
all to $O(\lambda^2)$ accuracy.

We calculate
$r^{(0)}$, the derivatives $r^{(0)}_{,\lambda}$, $r^{(0)}_{,A}$ and
$r^{(0)}_{,u}$ and the mixed derivatives $r^{(0)}_{,\lambda \lambda}$,
$r^{(0)}_{,\lambda A}$ and $r^{(0)}_{,\lambda u}$ at the points 
$(Q,P)$ on the world-tube by fourth order interpolation.

The null metric variables are then obtained  by
fourth order interpolation at the Bondi coordinate points $(r,Q,P,u)$, 
corresponding to the Cartesian coordinate points $(x^i,T)$

\subsection{Reversing the coordinate transformation}

The following describes the procedure to obtain the {\it contravariant} null
metric $\tilde{\eta}^{\tilde{\alpha}\tilde{\beta}}$ in affine coordinates
$\tilde{y}^{\alpha}=(u,q,p,\lambda)$ on points on the characteristic
grid. This is 90\% of the work in obtaining the Cartesian metric at
those points. What remains is to contract this null metric with the
Jacobian of the coordinate transformation
$x^{\mu}=x^{\mu}(\tilde{y}^{\alpha})$. The Jacobian is given
in Eq.~(\ref{eq:jacob}) as a series expansion on the parameter $\lambda$.
We will use this expression, and consequently we need to work with the 
contravariant (rather than the covariant) metric components,
since they will transform with the already known Jacobian as:
\begin{equation}
  g^{\mu\nu}  =
   {{\partial x^{\mu}} \over {\partial \tilde{y}^{\alpha}}}
   {{\partial x^{\nu}} \over {\partial \tilde{y}^{\beta}}}
    \tilde{\eta}^{\tilde{\alpha}\tilde{\beta}}.
	\label{eq:transform}
\end{equation}

\subsubsection{The metric on the sphere}

Recall that on each stereographic patch we introduce complex null
vectors on the sphere $q^{A}=(P/2)(\delta^{A}_{2}+i\delta^{A}_{3})$ 
where $P=1+\xi\bar\xi$ and the $q^{A}$ satisfy the orthogonality condition
$\bar{q}_{A}q^{A}=2$. The $q^{A}$ define a metric on the sphere 
$q_{AB} = (4/P^{2})\delta_{AB}$ and 
$q_{A}=(2/P)(\delta_{A}^{2}+i\delta_{A}^{3})=q_{AB} q^{B}$.

By construction, the null metric restricted to the sphere, $\eta_{AB}$,
is the same in Bondi coordinates $y^\alpha$ and affine coordinates
$y^{\tilde{\alpha}}$. It can be expanded as:

\begin{equation}
  \tilde{\eta}_{\tilde{A}\tilde{B}}   \equiv \eta_{AB}
   = \alpha q_{A} q_{B} + \bar{\alpha} \bar{q}_{A} \bar{q}_{B}
   + \delta q_{A} \bar{q}_{B} + \bar{\delta} \bar{q}_{A} q_{B}
\end{equation}
The values of $\alpha$ and $\delta$ follow from the orthogonality condition
\begin{equation}
   4\, \alpha = \eta_{AB} \bar{q}^{A} \bar{q}^{B}, \quad
   4\, \delta  = \eta_{AB} \bar{q}^{A} q^{B} = 4\, \bar{\delta},
\end{equation}
From the definition of $J$ and $K$
\begin{equation}
   J=\frac{1}{2} q^{A} q^{B} h_{AB}, \quad
   K=\frac{1}{2} q^{A} \bar q^{B} h_{AB}, \quad
\end{equation}
and $\eta_{AB} = r^2 h_{AB}$, it follows that
\begin{equation}
   \alpha = \frac{1}{2} r^{2} \bar{J}, \quad
   \delta  = \frac{1}{2} r^{2} K,
\end{equation}
so we can write the metric on the sphere in terms of $J$ and $K$ as
\begin{equation}
   \eta_{AB} = \frac{1}{2} r^{2} \left( 
   J \bar{q}_{A} \bar{q}_{B} + \bar{J} q_{A} q_{B} 
   + K q_{A} \bar{q}_{B} + K \bar{q}_{A} q_{B} \right)
\end{equation}

From the expressions for the $q^{A}$, the components of the sphere metric
are given by:
\begin{equation}
   \eta_{qq} = \frac{4\, r^{2}}{P^{2}} \left( K + \Re(J) \right), \quad
   \eta_{pp} = \frac{4\, r^{2}}{P^{2}} \left( K - \Re(J) \right), \quad
   \eta_{qp} = \frac{4\, r^{2}}{P^{2}} \Im(J)
\end{equation}

Notice that given the Bondi coordinates of a point, the determinant is known
\begin{equation}
   \det(\eta_{AB}) = r^{4} \det(q_{AB}) = 16\, r^{4} / P^4,
\end{equation}
therefore the inverse sphere metric ${\eta}^{AB}$ such that 
${\eta}^{AB} {\eta}_{BC} = \delta^{A}_{C}$
has coefficients
\begin{eqnarray}
   \eta^{qq} &=& \tilde{\eta}^{\tilde{q}\tilde{q}} = \frac{P^{2}}{4\, r^{2}} \left( K - \Re(J) \right), \nonumber \\
   \eta^{pp} &=& \tilde{\eta}^{\tilde{p}\tilde{p}} = \frac{P^{2}}{4\, r^{2}} \left( K + \Re(J) \right), \nonumber \\
   \eta^{qp} &=& \tilde{\eta}^{\tilde{q}\tilde{p}} = - \frac{P^{2}}{4\, r^{2}} \Im(J).
\end{eqnarray}

\subsubsection{The radial-angular metric coefficients}

We can write the metric coefficients in terms of a single coefficient $\gamma$
\begin{equation}
   \tilde{\eta}^{\lambda \tilde{A}} = \bar\gamma q^{\tilde{A}} + \gamma \bar{q}^{\tilde{A}}
\end{equation}
In components
\begin{equation}
   \tilde{\eta}^{\lambda p} = P \Re(\gamma), \quad
   \tilde{\eta}^{\lambda q} = P \Im(\gamma)
\end{equation}

The value of $\gamma$ follows from the expression for the metric function $U$
\begin{equation}
   U = U^{A} q_{A} = \frac{\eta^{rA}}{{\eta^{ru}}} q_{A}
     = - \left(\tilde{\eta}^{\lambda \tilde{A}}
       + \frac{r_{,\tilde{B}}}{r_{,\lambda}} \tilde{\eta}^{\tilde{A}\tilde{B}} \right) q_{\tilde{A}} ,
\end{equation}
\begin{equation}
   2\, \gamma = - U 
   - \frac{r_{,\tilde{B}}}{r_{,\lambda}} \tilde{\eta}^{\tilde{A}\tilde{B}} q_{\tilde{A}}
\end{equation}
Recall that the derivatives $r_{,A}$, $r_{,u}$ are at constant $\lambda$
and $r_{,\lambda}$ can be read off directly from the Bondi metric function 
$\beta$
\begin{equation}
   r_{,\lambda} = e^{-2\, \beta}
\end{equation}

From the expressions for $\tilde{\eta}^{\tilde{A}\tilde{B}}$ and $q_{\tilde{A}}$,
\begin{eqnarray}
   r_{,\tilde{B}} \tilde{\eta}^{\tilde{A}\tilde{B}} q_{\tilde{A}} &=& \frac{2}{P} \left[
  r_{,2} \left(\tilde{\eta}^{pp} + i \tilde{\eta}^{pq} \right) 
+ r_{,3} \left(\tilde{\eta}^{pq} + i \tilde{\eta}^{qq} \right) \right] \nonumber \\
  &=& \frac{P}{2\, r^{2}} \left[ r_{,2} \left(K - J \right)
     + i r_{,3} \left(K + J \right) \right]
\end{eqnarray}
thus the complex coefficient sought is given by
\begin{equation}
   \gamma = - \frac{1}{2} U  - e^{2\, \beta}
   \frac{P}{4\, r^{2}} \left[ r_{,2} \left(K - J \right)
     + i r_{,3} \left(K + J \right) \right]
\end{equation}
The $r_{,A}$ can be obtained at the point where the coordinate transformation 
is being performed to $O(\lambda^{2})$, by using the angular derivatives of 
$r_{,\lambda}$.  To this end, note that for fixed (q,p):
\begin{equation}
   r_{,A} = r^{(0)}_{,A} + r_{,\lambda A} \lambda
\end{equation}
where the derivatives on the right hand side are computed on the world-tube.

\subsubsection{The $\tilde{\eta}^{\lambda \lambda}$ metric function}

By reversing the procedure to obtain $W$, i.e. from 

\begin{equation}
   W = \frac{1}{r} \left( r_{,\lambda} \tilde{\eta}^{\lambda\lambda}
   + 2 \, \left( r_{,\tilde{A}}\,\tilde{\eta}^{\lambda \tilde{A}} - r_{,u} \right)
   + \frac{r_{,\tilde{A}}\,r_{,\tilde{B}}}{r_{,\lambda}}\, \tilde{\eta}^{\tilde{A}\tilde{B}} - 1 \right)
\end{equation}
we obtain the remaining component of the null metric 

\begin{equation}
   \tilde{\eta}^{\lambda\lambda} = e^{2\, \beta} \left( r W + 1 
   - 2 \, \left( r_{,\tilde{A}}\,\tilde{\eta}^{\lambda \tilde{A}} - r_{,u} \right)
   - e^{2\, \beta}\, r_{,\tilde{A}}\,r_{,\tilde{B}}\, \tilde{\eta}^{\tilde{A}\tilde{B}} \right)
\end{equation}
We have made use of the relation $r_{,\lambda} = e^{-2\, \beta}$.
The $r_{,u}$ and $r_{,\tilde{A}}$ are at constant $\lambda$, the former
can be evaluated to $O(\lambda)$ accuracy by using the value at
the world-tube. To get $r_{,u}$ to $O(\lambda^2)$, we need to calculate
$r_{,\lambda u} = -2\, e^{-2\,\beta} \beta_{,u}$ on the world-tube, and
use $r_{,u} = r^{(0)}_{,u} + r_{,\lambda u} \lambda$.

This completes the calculation of the final contravariant metric
component, and so the injection process may  be carried out at last, after
the final multiplication by the Jacobian described in
Eq.~(\ref{eq:transform}).

\subsubsection{Status of the algorithm}

The CCM algorithm described in the previous sections is quite new. It
has not yet been translated into a stable numerical scheme for black
hole collisions and interactions in 3-D. However, there has already been
extensive testing of these techniques in simpler situations. Recently,
significant codes have been built which implement these ideas for scalar
waves without gravity but with full 3-D Cartesian and null spherical
coordinate grid patches in flat space~\cite{flat1,flat2}. Codes have
also been written for 1-D problems with gravitation, both for spherical
self-gravitating scalar waves collapsing and forming black
holes~\cite{KG,excising} and for cylindrically symmetric vacuum
geometry~\cite{cylinder1,cylinder2}. What has not yet been accomplished
is the development of stable codes with both 3-D geometry and non-trivial
strong gravitation, although codes currently under development
demonstrate the stable evolution of a single moving black hole. Thus,
we may expect exciting progress in the near future, and the mysterious
details of collision between black holes may soon be unveiled.

\subsection{Acknowledgements} 

This work has been supported by the Binary
black Hole Grand Challenge Alliance, NSF PHY/ASC 9318152 (ARPA
supplemented) and by NSF PHY 9510895 and NSF INT 9515257 to the
University of Pittsburgh. N.T.B. thanks the Foundation for Research
Development, South Africa, for financial support. N.T.B. and R.A.I. thank
the University of Pittsburgh for hospitality. J.W. and R.A.I. thank the
Universities of South Africa and of Durban-Westville for their
hospitality.


\end{document}